\documentclass[aps,prl,reprint,superscriptaddress]{revtex4-2}

\usepackage{graphicx}
\usepackage{amsmath,amssymb}
\usepackage{hyperref}
\usepackage{siunitx}
\providecommand{\doi}[1]{\href{https://doi.org/#1}{doi:\,#1}}
\newcommand{\pt}{p_{\perp}}
\newcommand{\RAA}{R_{\mathrm{AA}}}
\newcommand{\vTwo}{v_{2}}
\newcommand{\etas}{\eta/s}
\newcommand{\tauz}{\tau_{0}}

\begin{document}

\title{Joint Bayesian analysis of soft and high-$p_\perp$ probes yields tighter constraints on QGP properties}

\author{Marko Djordjevic}
\affiliation{Bioinformatics Group, Faculty of Biology, University of Belgrade, Belgrade 11000, Serbia}

\author{Dusan Zigic}
\affiliation{Institute of Physics Belgrade, University of Belgrade, Belgrade 11080, Serbia}

\author{Igor Salom}
\affiliation{Institute of Physics Belgrade, University of Belgrade, Belgrade 11080, Serbia}

\author{Magdalena Djordjevic}
\email{magda@ipb.ac.rs}
\affiliation{Institute of Physics Belgrade, University of Belgrade, Belgrade 11080, Serbia}
\affiliation{Serbian Academy of Sciences and Arts, Belgrade 11000, Serbia}

\begin{abstract}
%
To extract bulk QGP properties, we perform a joint Bayesian calibration of bulk-medium parameters using low-$\pt$ bulk and high-$\pt$ tomography within a common medium evolution. Low-$\pt$ observables are computed with \textsc{TRENTo}+\textsc{VISHNU}; temperature profiles are passed to \textsc{DREENA-A} to predict light/heavy $R_{\mathrm{AA}}(\pt)$ and $v_2(\pt)$. Gaussian-process emulation enables Hamiltonian Monte Carlo sampling of the low-$\pt$-only and joint posteriors. The low-$\pt$-only case underpredicts high-$\pt$ anisotropy; the joint calibration matches both sectors and markedly tightens bulk-parameter constraints, demonstrating the added power of high-$\pt$ data.

\end{abstract}
\maketitle

\section{Introduction}
Relativistic heavy-ion collisions at RHIC and the LHC create short-lived droplets of deconfined QCD matter, the quark--gluon plasma (QGP), favoring a small specific shear viscosity $\eta/s$ near the QCD transition~\cite{HeinzSnellings2013,SchaeferTeaney2009,KSS2005,Lacey2006,Lisa2011,HeinzSchenke2024Review}. Over the past decade, Bayesian parameter estimation has become a standard tool for quantitative model-to-data comparison in the soft sector, enabling simultaneous constraints on initial conditions and bulk-evolution parameters while propagating experimental and modeling uncertainties~\cite{Bernhard2016,BernhardNaturePhys2019,Auvinen2020,NijsTrajectum2021,JETSCAPEsoft2021,Parkkila2021} (see Ref.~\cite{Paquet2024Review} for a recent overview).
While these analyses have substantially improved quantitative constraints, some bulk-medium properties remain difficult to determine precisely from low-$\pt$ data alone: soft-sector observables can have limited sensitivity to transport properties away from the transition region and/or to the earliest, hottest stage of the evolution~\cite{Nagle2011,Niemi2011,Plumari2015}, and different parameter combinations can still describe the low-$\pt$ data comparably well, with additional sensitivities to modeling choices and external constraints~\cite{NijsVandSchee2022,Heffernan2024}.

High-$\pt$ probes provide complementary information through QGP tomography: partons produced in early hard scatterings lose energy while traversing the medium, leading to jet-quenching observables such as the nuclear modification factor $R_{\mathrm{AA}}(\pt)$ and high-$\pt$ anisotropies $v_n(\pt)$, which are sensitive to the space--time temperature profile sampled along the parton path~\cite{MajumderVanLeeuwen2011,Connors2018,CaoJetReview2024,Zigic:2021rku,Karmakar:2024jak}.
Heavy flavor adds mass-dependent sensitivity through the dead-cone effect (see e.g.,~\cite{Djordjevic:2016vfo,Zigic:2021rku}), and the precision and breadth of high-$\pt$ measurements continue to grow in the high-luminosity heavy-ion programs at the LHC and at RHIC (sPHENIX)~\cite{CitronYellowReport2019,BelmontsPHENIX2024}.
Only a handful of exploratory Bayesian studies have addressed hard probes, mainly constraining effective jet-transport or heavy-flavor transport parameters from selected high-$p_\perp$ observables~\cite{He:2018gks,JETSCAPEqhat2021,EhlersJETSCAPE2024,XieKeWang2023,FanMetric2024,DuJETSCAPE2025,XueHeavyQBayes2025}. Here we take the next step by developing a proof-of-concept framework for a joint soft+hard calibration within a single hydrodynamic background, simultaneously constraining the bulk-medium evolution parameters with light- and heavy-flavor $R_{AA}(p_\perp)$ and $v_2(p_\perp)$.

In this Letter we couple a soft-sector medium evolution (\textsc{TRENTo}+\textsc{VISHNU}\cite{MorelandTRENTo2015,Bernhard2016,VISHNU}) to the dynamical energy-loss framework \textsc{DREENA-A}~\cite{Zigic:2021rku}, enabling low-$\pt$ bulk data and high-$\pt$ light- and heavy-flavor $R_{\mathrm{AA}}(\pt)$ and $v_2(\pt)$ to constrain the same bulk-medium parameters within a common background.
We contrast a low-$\pt$-only calibration with a joint soft+hard calibration to test whether a bulk-constrained medium evolution can simultaneously describe high-$\pt$ suppression and anisotropy, and to quantify how hard-probe information affects the inferred parameter region.

\section{Methodology}
\noindent\textit{Medium model and parameter design.}
Initial entropy density profiles are generated with the parametric initial condition model \textsc{TRENTo}~\cite{MorelandTRENTo2015}, and the subsequent QGP evolution is simulated with $(2+1)$-dimensional viscous hydrodynamics using \textsc{VISHNU}~\cite{Bernhard2016,VISHNU}. 
We assume \textsc{TRENTo} with $p=0$ and do not include an explicit pre-equilibrium free-streaming stage, as we showed that free streaming is disfavored by high-$\pt$ data~\cite{Stojku:2020wkh}. 
We use an averaged-medium setup: for each centrality class, initial conditions are averaged with reaction planes aligned and evolved in a single hydrodynamic run to obtain the temperature profiles used in the analysis. 
For clarity and computational tractability in this first study, we restrict ourselves to a three-parameter space with a constant $\etas$, varying (i) the overall normalization \texttt{norm} (60--360), (ii) the starting time $\tauz \in [0.2,1.3]~\mathrm{fm}$, and (iii) $\etas \in [0.02,0.2]$, while fixing all other parameters to Ref.~\cite{Karmakar2023PRC108}. 

\noindent\textit{Low-$\pt$ and high-$\pt$ observables.}
For each hydrodynamic simulation we compute low-$\pt$ bulk observables in four centrality classes (10--20\%, 20--30\%, 30--40\%, and 40--50\% Pb+Pb at 5.02~TeV): identified-hadron rapidity densities $\mathrm{d}N/\mathrm{d}y$ and mean transverse momenta $\langle \pt \rangle$ for $\pi$, $K$, and $p$, and a $\pt$-integrated $h^\pm$ elliptic flow $\langle \vTwo \rangle$ (averaged over $\pt<2$~GeV). 
The same temperature profiles are passed to the \textsc{DREENA-A} framework~\cite{Zigic:2021rku,Zigic:2022xks} to compute high-$\pt$ $\RAA(\pt)$ and $\vTwo(\pt)$ for inclusive charged hadrons $h^\pm$ (same centrality as for low-$\pt$) and for $D^0$ mesons (30--50\%). 
We use one representative dataset per observable: ALICE for bulk~\cite{ALICEv2,ALICEMult} and $D^0$ $\RAA$~\cite{ALICEDRAA}, ATLAS for $h^\pm$ $\RAA$ and $\vTwo$~\cite{ATLASRAA,ATLASv2}, and CMS for $D^0$ $\vTwo$~\cite{CMSDv2}; a full multi-experiment analysis with correlated systematics is left for future work.

\noindent\textit{Principal-component reduction, emulation and Bayesian inference.}
The coupled model is evaluated on a Latin-hypercube design of $\sim200$ points in the three-dimensional parameter space $\boldsymbol{\theta}=(\texttt{norm},\tauz,\etas)$. To reduce the dimensionality of the multi-observable output vector and mitigate strong correlations among its components, we standardize each observable across the design and perform principal-component (PC) decompositions separately for the low-$\pt$ and high-$\pt$ sectors, retaining the minimum number of PCs needed to capture at least $99\%$ of the total variance. For low-$\pt$, identified-hadron yields and mean transverse momenta are reduced jointly, yielding two PCs for the combined $\{dN/dy,\langle\pt\rangle\}$ set, while the four centrality values of the integrated elliptic flow $\langle\vTwo\rangle$ are well described by a single dominant PC. For high-$\pt$ we decompose the combined $\{1-\RAA(\pt),\,\vTwo(\pt)\}$ vector for $h^\pm$ and $D^0$, exploiting the strong correlation between suppression and anisotropy~\cite{StojkuAnisotropy2022}; three PCs are retained.

Independent Gaussian-process (GP) emulators are trained for each retained PC coefficient as a function of $\boldsymbol{\theta}$ and used as a fast surrogate in two calibrations: (i) low-$\pt$ PCs only and (ii) combined low-$\pt$+high-$\pt$ PCs. Emulator accuracy is validated on held-out design points, yielding out-of-sample $R^2>0.99$ for all retained PCs. Experimental measurements are projected into the same PC space using the same standardization and PC basis learned from the design, and their statistical and systematic uncertainties are propagated by the linear transformation. We assume uniform priors within the parameter ranges specified above and construct a Gaussian likelihood in PC space with diagonal covariance; for each PC coefficient, the total uncertainty combines propagated experimental errors with an emulator-uncertainty estimate (GP predictive variance and a cross-validation term) added in quadrature. We sample the posterior with Hamiltonian Monte Carlo using finite-difference gradients of the log posterior. Posterior predictive distributions are obtained by evaluating the emulators on posterior samples and transforming back to the experimental basis via the inverse PC and standardization maps (and $1-\RAA\rightarrow\RAA$).

\section{Results and discussion}
We compare two Bayesian calibrations of the 
vector $\boldsymbol{\theta}$:
(i) an inference constrained only by low-$\pt$ (soft-sector) observables, followed by an out-of-sample test on high-$\pt$ (hard-probe) observables computed with \textsc{DREENA-A}; and
(ii) a joint inference constrained simultaneously by the combined low-$\pt$ and high-$\pt$ datasets.
The goal is to test whether a parameter set tuned to bulk observables can also describe high-$\pt$ suppression and anisotropy, and to quantify the additional constraining power gained by including hard-probe information in the likelihood.

\noindent\textit{Low-$\pt$ inference and out-of-sample hard-probe test.}
\begin{figure}[h]
    \centering
    \includegraphics[width=0.45\textwidth]{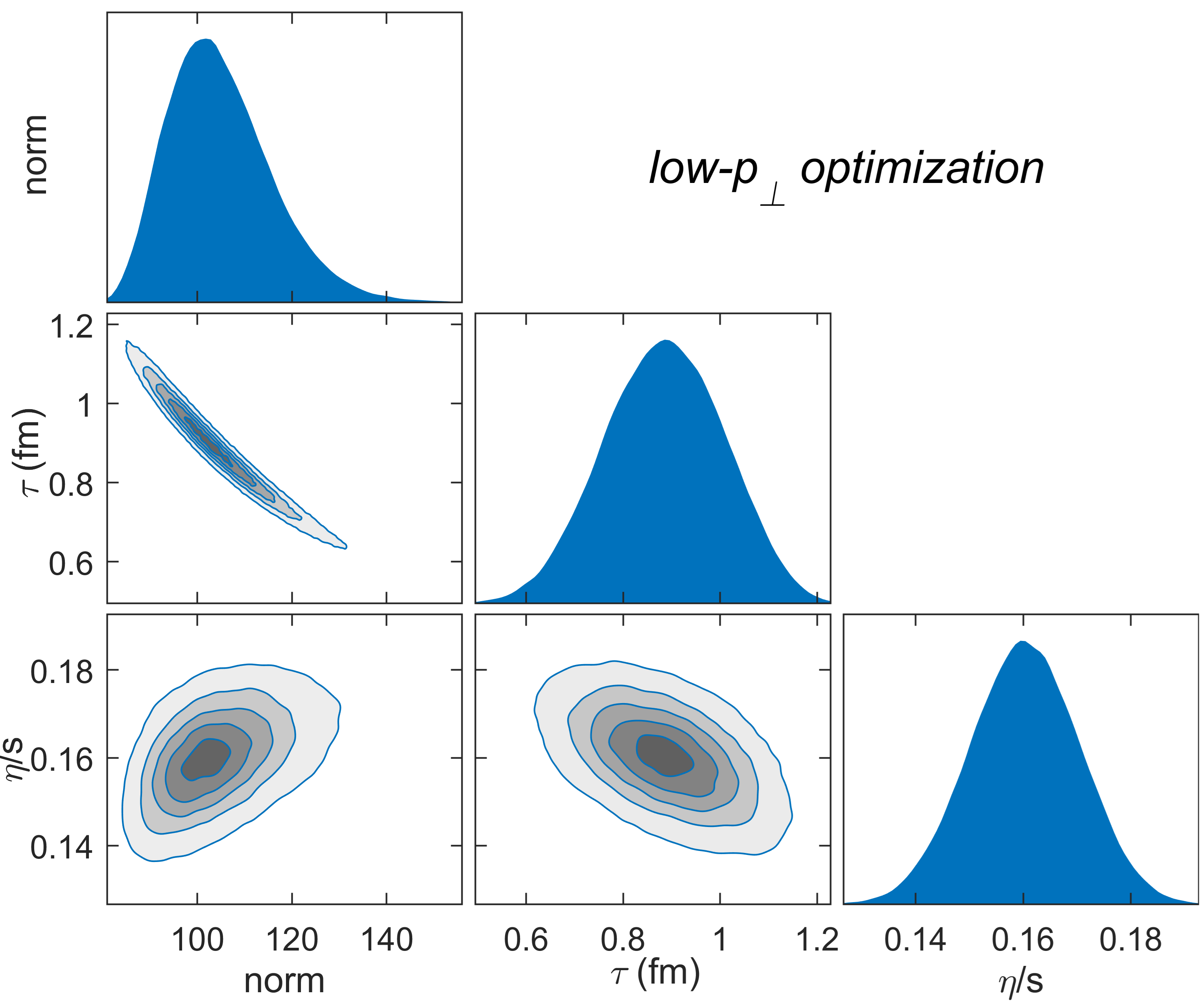}
    \caption{Posterior distributions of the calibration parameters ($\tau_0$, $\eta/s$, and \texttt{norm}) from the low-$\pt$-only Bayesian inference. Diagonal panels show marginalized 1D posteriors; off-diagonal panels show 2D joint contours.}
    \label{fig:corner_lowpt}
\end{figure}
Figure~\ref{fig:corner_lowpt} shows the HMC posterior obtained from the low-$\pt$ dataset.
The distribution exhibits a clear anti-correlation between norm and $\tau_0$, reflecting the partial compensation between the overall entropy normalization and the hydrodynamic starting time in bulk observables (see, e.g., Ref.~\cite{Virta:2025}).
The posterior for $\etas$ is also constrained by the low-$\pt$ data, but the remaining width of the marginal and joint distributions indicates that sizable uncertainties persist after a soft-only calibration.

\begin{figure}[h]
  \centering
  \includegraphics[width=0.48\textwidth]{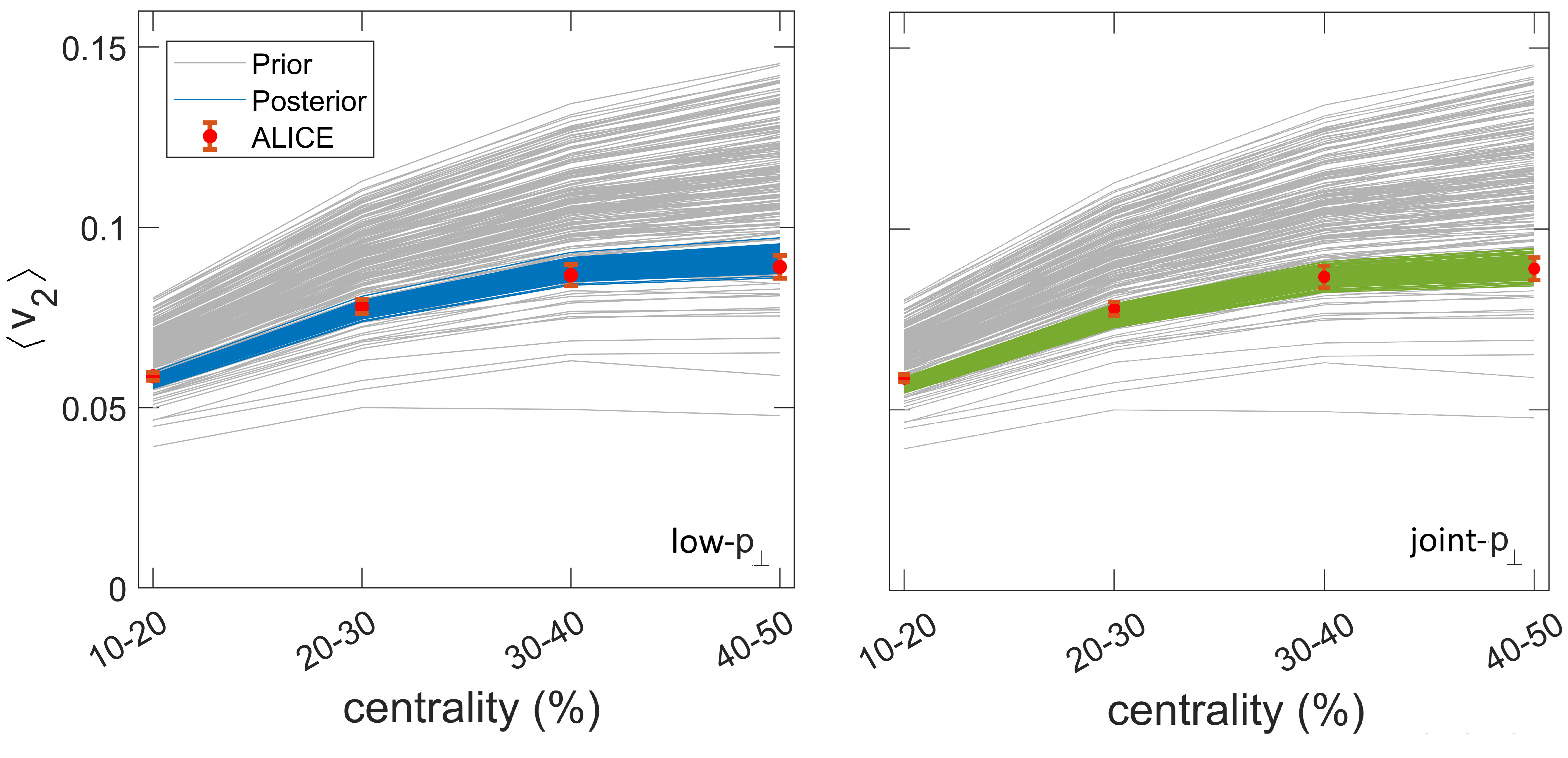}
\caption{Prior (gray) and posterior predictions for the $p_{\mathrm{T}}$-integrated $h^\pm$ elliptic flow $\langle v_2\rangle$ versus centrality: \emph{low-$\pt$ only} (left, blue) and \emph{joint} (low-$\pt$+high-$\pt$, right, green). Red points denote the ALICE measurements~\cite{ALICEv2} used in the calibration.}
  \label{fig:lowpt_v2}
\end{figure}

Propagating posterior samples from the low-$\pt$ calibration to the soft observables, the model reproduces identified-hadron multiplicities and  $\langle \pt \rangle$  (not shown), and Fig.~\ref{fig:lowpt_v2} (left) shows the corresponding prior-to-posterior narrowing and agreement for $\langle v_2\rangle$ versus centrality.

\begin{figure}[h]
  \centering
  \includegraphics[width=0.48\textwidth]{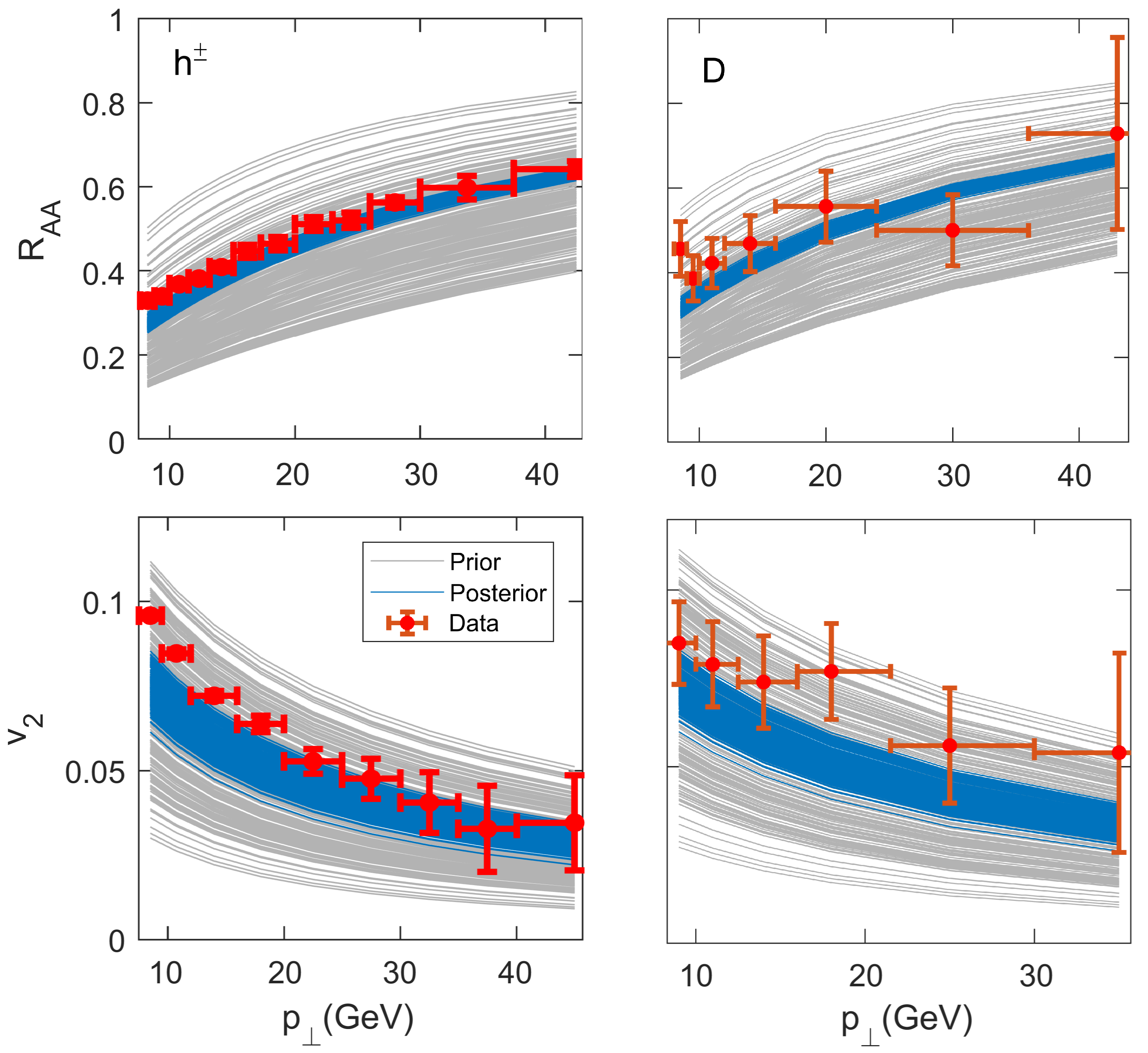}
\caption{High-$\pt$ out-of-sample test of the low-$\pt$ only posterior propagated through \textsc{DREENA-A}. Shown are $\RAA$ and $\vTwo$ vs.~$\pt$ for $h^\pm$ (left, 30--40\%) and $D^0$ (right, 30--50\%). Prior (gray) and posterior (blue) bands are compared to ATLAS ($h^\pm$)~\cite{ATLASRAA,ATLASv2}, ALICE ($D^0$ $\RAA$)~\cite{ALICEDRAA}, and CMS ($D^0$ $\vTwo$)~\cite{CMSDv2} data (red). $h^\pm$ results were computed for all four centralities; only 30--40\% is shown for brevity.}
  \label{fig:highpt_chD_lowptpost}
\end{figure}

Propagating the low-$p_\perp$ posterior through \textsc{DREENA-A} and comparing to the high-$p_\perp$ data not included in the likelihood (Fig.~\ref{fig:highpt_chD_lowptpost}), the soft-only calibration remains broadly compatible with $R_{AA}(p_\perp)$ but systematically underpredicts $v_2(p_\perp)$ for both light and heavy flavor.
This behavior is consistent with the long-standing difficulty of simultaneously describing high-$p_\perp$ $R_{AA}$ and azimuthal anisotropy within energy-loss–based approaches (see e.g.,~\cite{ZhaoRAAv2PRL2022,Stojku:2020wkh}).
Since $\RAA(\pt)$ primarily constrains the overall quenching strength whereas $\vTwo(\pt)$ constrains its path-length- and geometry-driven modulation, this indicates that the soft-sector data alone do not provide sufficient constraining power to determine a medium evolution that simultaneously reproduces the observed high-$\pt$ anisotropy, motivating the inclusion of high-$\pt$ observables in the Bayesian calibration.

\noindent\textit{Joint low-$\pt$+high-$\pt$ inference.}
\begin{figure}[h]
  \centering
  \includegraphics[width=0.49\textwidth]{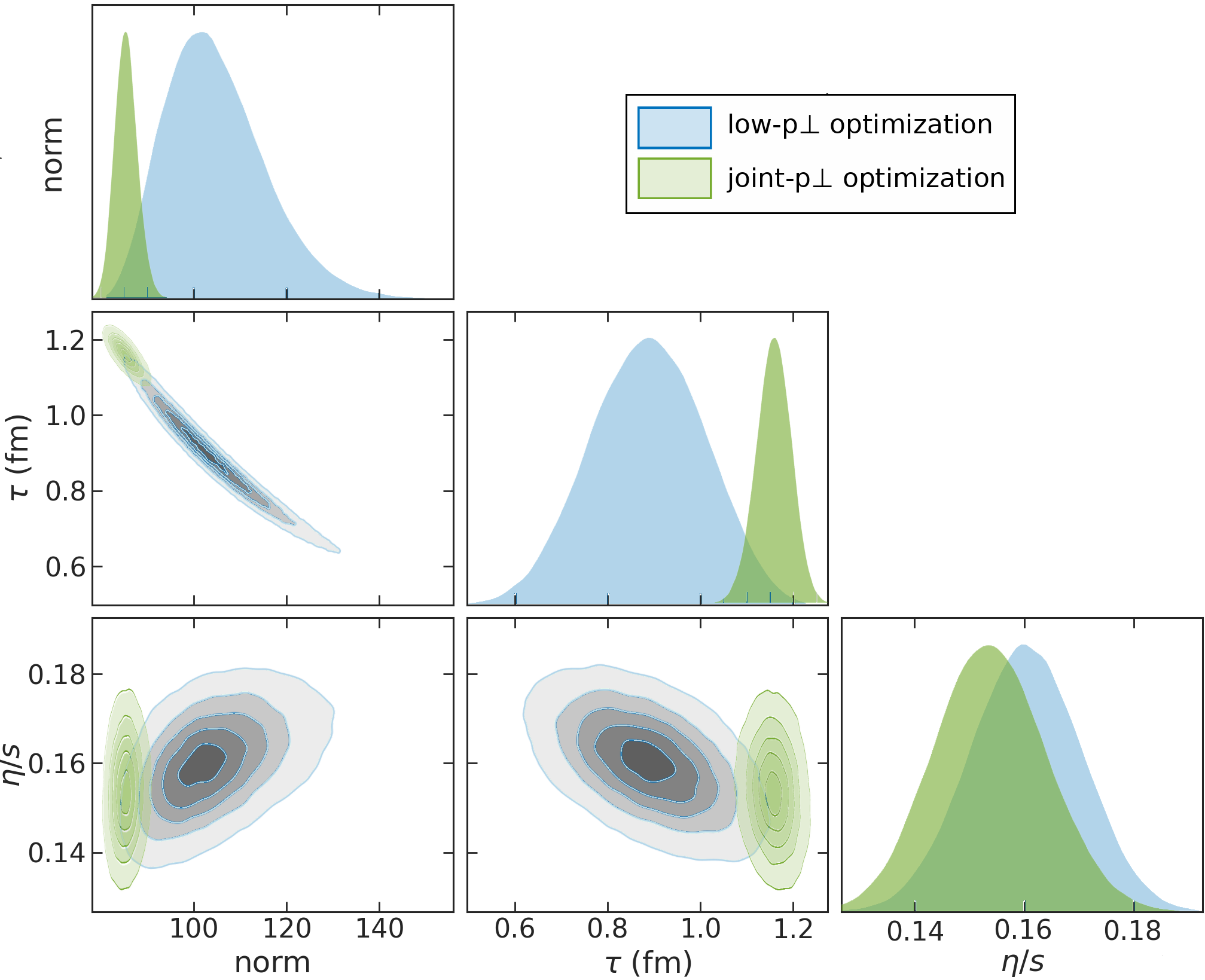}
  \caption{Same as Fig.~\ref{fig:corner_lowpt}, but comparing overlaid posterior distributions from the low-$p_\perp$-only (blue) and joint low-$p_\perp$+high-$p_\perp$ (green) calibrations.}
  \label{fig:corner_joint}
\end{figure}
Including the high-$\pt$ dataset in the likelihood and repeating the inference yields the joint posterior shown in Fig.~\ref{fig:corner_joint} (green), overlaid with the low-$\pt$-only result (blue).
Relative to the low-$\pt$-only case, the joint posterior is substantially narrower and shifts toward smaller \texttt{norm} and larger $\tauz$, while $\etas$ changes more moderately, indicating that hard-probe data lift degeneracies present in a soft-only analysis.

The combined analysis does not degrade the description of low-$\pt$ observables: the joint posterior agrees with $\langle v_2\rangle$ (Fig.~\ref{fig:lowpt_v2}, right), and the identified-hadron dN/dy and $\langle p_\perp\rangle$ observables are also reproduced within uncertainties (not shown). Thus, hard-probe information narrows the set of medium evolutions compatible with both sectors without compromising the bulk description.

\begin{figure}[h]
  \centering
  \includegraphics[width=0.48\textwidth]{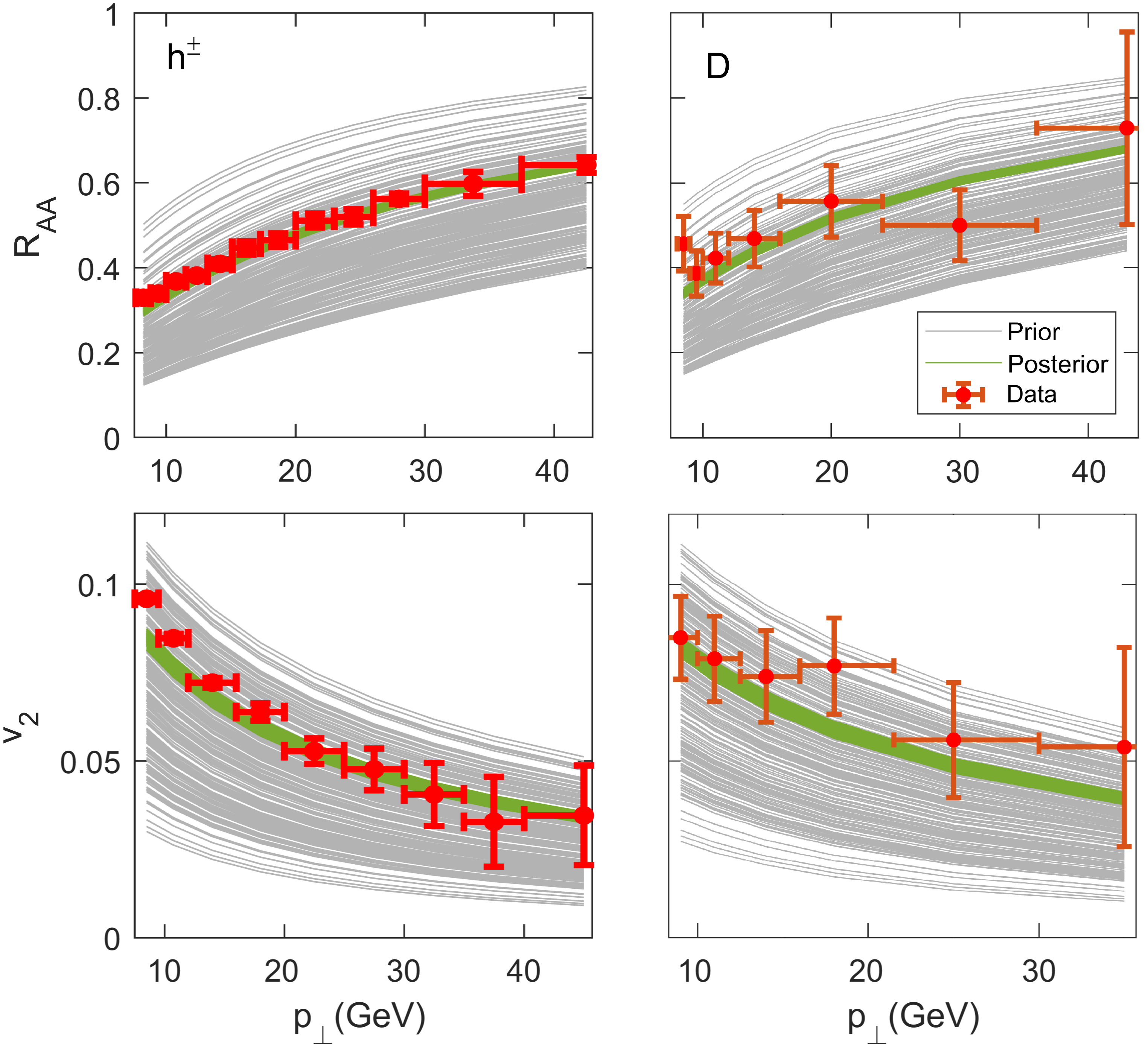}
  \caption{Same as Fig.~\ref{fig:highpt_chD_lowptpost}, but using posterior samples from the joint calibration (green).}
  \label{fig:highpt_chD_joint}
\end{figure}

Most importantly, the joint posterior resolves the high-$p_\perp$ anisotropy deficit: Fig.~\ref{fig:highpt_chD_joint} shows good agreement with both $R_{AA}(p_\perp)$ and $v_2(p_\perp)$ for $h^\pm$ and simultaneously describes $D$-meson suppression and elliptic flow.
Compared to Fig.~\ref{fig:highpt_chD_lowptpost}, the improvement is driven primarily by $\vTwo(\pt)$, demonstrating that a simultaneous quantitative description of low-$\pt$ bulk observables and high-$\pt$ light- and heavy-flavor observables is achievable once the calibration is constrained by both sectors.

At the same time, the joint posterior selects a subset of the soft-only degeneracy ridge, indicating no evidence (within the current modeling assumptions) that soft- and hard-sector data require mutually incompatible medium parameters.
Rather, the hard-probe data sharpen the extraction by providing complementary constraints that are most directly encoded in high-$\pt$ anisotropies $\vTwo(\pt)$ for both light and heavy flavor.
Overall, the results quantify the central message suggested by the prior-vs-posterior comparisons: in this framework, high-$\pt$ observables act as an additional, independent tomographic constraint that reduces low-$\pt$-only degeneracies and strengthens bulk-parameter determination.

\section{Conclusion}
We presented a proof-of-concept joint Bayesian inference that combines low-$\pt$ bulk observables with high-$\pt$ tomography observables within a common medium evolution.
The calibration is performed in a reduced principal-component basis (three PCs per sector, capturing 99\% of the variance) and mapped back to experimental observables for posterior predictive comparisons.
Comparing a low-$\pt$-only calibration to a joint low-$\pt$+high-$\pt$ calibration, we find that the soft-only posterior reproduces the bulk data and remains broadly compatible with $\RAA(\pt)$, but systematically underestimates high-$\pt$ $\vTwo(\pt)$; including high-$\pt$ observables resolves this anisotropy deficit and substantially narrows the inferred parameter posteriors.
Given the strong correlation between light-flavor high-$\pt$ $\RAA$ and $\vTwo$, heavy-flavor suppression and anisotropy add a crucial, more weakly correlated constraint—treated on the same footing within our DREENA framework for both light and heavy flavors.

Looking ahead, high-precision LHC/RHIC programs will enable additional weakly correlated constraints, especially from $D$ and $B$-meson high-$\pt$ $\RAA$, $\vTwo$, and higher harmonics~\cite{CitronYellowReport2019,BelmontsPHENIX2024}. 
Given the low effective dimensionality observed in both soft and hard datasets, further progress will rely on extending the calibration to additional heavy-flavor observables, reconstructed jets (including their $R$ dependence), higher flow harmonics, and complementary systems and energies (e.g.\ RHIC and lighter-ion runs at the LHC~\cite{PablosTakacsOO2025}), as well as combining multiple experiments with correlated uncertainties. Although reconstructed-jet observables are not included here, a simultaneous description of high-$\pt$ hadron and jet suppression provides a stringent consistency test~\cite{CasalderreySolana2019}, while the jet-cone radius ($R$) dependence adds sensitivity to out-of-cone energy redistribution and medium response~\cite{MehtarTaniPablosTywoniuk2021,BarretoJEWEL2025}; our cone-dependent radiative and collisional energy loss extensions~\cite{KarmakarDjordjevicJetRad,DIDJetColl2026} provide the ingredients to incorporate jets into DREENA framework.
Methodologically, next steps include expanding the inferred parameter space beyond the present three-parameter setup, incorporating event-by-event fluctuations and fluctuation-sensitive observables, and exploiting the joint low-$\pt$+high-$\pt$ information content of forthcoming datasets, in line with broader efforts to incorporate hard probes into Bayesian analyses~\cite{JETSCAPEsoft2021}.

\section*{Acknowledgments} This work is supported by the Ministry of Science and Technological Development of the Republic of Serbia and by the Serbian Academy of Sciences and Arts.

\end{document}